\documentclass[onecolumn,authoryear]{els-mrw} 

\usepackage{amsmath,amssymb,amsfonts,amsthm,makeidx,graphicx}
\usepackage{txfonts}
\usepackage{helvet}

\begin{document}

\chapter{Giant branch planetary systems: Dynamical and radiative evolution}\label{chap1}

\author[1]{Alexander J. Mustill}%
\address[1]{\orgname{Lund University}, \orgdiv{Department of Physics}, \orgaddress{Box 118, SE-22 100 Lund, Sweden}}

\articletag{XXX Chapter Article tagline: update of previous edition,, reprint..}

\maketitle

\begin{glossary}[Glossary]

\term{Adiabatic} [Dynamics] Slow in relation to some characteristic timescale.
 (usually the orbital timescale in this context).
 
\term{Asymptotic Giant Branch} Phase in a star's life following exhausting of helium fuel in the core, and preceding the WD.

\term{Main Sequence} Stage of a star's life where it is fusing hydrogen in its core.

\term{Planetary Nebula} The expelled stellar envelope after the AGB, surrounding and illuminated by the star as it transitions to a WD. Nothing to do with planets.

\term{Planetesimal} Generic term for a small body, asteroid or comet.

\term{Radiation/radiative forces} Forces acting on orbiting bodies through absorption, emission and scattering of radiation, including radiation pressure, Poynting--Robertson drag, the Yarkovsky effect, and the YORP effect. 

\term{Red Giant Branch} Evolutionary state after the Main Sequence, when hydrogen fuel in the core is exhausted but prior to the ignition of helium fuel.

\term{Tidal} Related to a \emph{differential} gravitational potential felt across an extended body such as a star or planet.

\term{White Dwarf} Final state of stellar evolution, composed of electron-degenerate matter with no nuclear burning.

\term{}

\end{glossary}

\begin{glossary}[Abbreviations]
\begin{tabular}{@{}lp{34pc}@{}}
AGB &Asymptotic Giant Branch\\
MMR &Mean motion resonance\\
MS &Main Sequence\\
PN &Planetary Nebula\\

PR &Poynting--Robertson\\
RGB &Red Giant Branch\\
WD &White Dwarf\\
YORP &Yarkovsky--O'Keefe--Radzievskii--Paddack\\
\end{tabular}
\end{glossary}

\begin{abstract}[Abstract]
In seven billion years, the Sun will be dead. As stars like the Sun pass from their present state to that of a dead white dwarf star, they undergo two phases of extremely high luminosity and radius---the red giant branch and the asymptotic giant branch---during which they will lose half or more of their mass. These changes to the star have a significant impact on orbiting planets, asteroids and comets. The large stellar radius (beyond the current orbit of the Earth) leads to the engulfment of bodies entering the stellar envelope, a process enhanced by strong tidal interactions. The high luminosity affects bodies' orbits and physical properties, while mass loss can later trigger the destabilisation of bodies around white dwarfs. It is necessary to understand these processes to understand both the future of our Solar System, and to interpret growing observations of planetary systems around evolved stars.
\end{abstract}

\section{Introduction}\label{sec:intro}

\begin{BoxTypeA}{Key points}
\begin{itemize}
    \item Stars have a finite life, limited by their nuclear fuel. When hydrogen is exhausted in the core, the Sun swells to a large size and luminosity as a \emph{red giant branch (RGB)} star. It will later begin fusing helium in its core, and when that is exhausted it will again swell to an even greater size as an \emph{asymptotic giant branch (AGB)} star. Following that, it will expel its envelope and become a ``dead'' \emph{white dwarf,} not undergoing any further nuclear reactions.
    \item The most significant changes happen on the AGB, where the Sun will grow to become larger than Earth's orbit, increase in brightness to several thousand times today's value, and lose half of its mass.
    \item The hugely increased stellar radius can engulf and destroy bodies such as planets, either directly, or by dragging them in through enhanced \emph{tidal} forces. The result of this should be a large cleared region round a white dwarf, an expectation increasingly at odds with observations.
    \item The extreme luminosities can change the orbits of small dust grains and asteroids up to several tens of km through \emph{radiation forces} such as \emph{radiation pressure, Poynting--Robertson drag,} and the \emph{Yarkovsky effect}. These bodies are also potentially subject to sublimation or loss of their volatiles. Planets that evade the reach of the star's strong tidal forces may have their atmospheres stripped or chemically altered by the extreme irradiation.
    \item The loss of mass from the star causes the orbits of any surviving bodies to expand. When more than one body is present in a system, forces between them become relatively stronger after the star has lost mass. This can result in orbital instabilities which may be responsible for explaining planetary phenomena close to white dwarfs, such as transiting planets and  planetesimals, discs of dusty debris and vaporised rock, and rocky material deposited into white dwarf atmospheres.
\end{itemize}
\end{BoxTypeA}

While stars like the Sun live for a long time, they do not live forever. Eventually they exhaust their source of nuclear fuel, and ultimately, unless sufficiently massive, end their lives as slowly cooling white dwarfs. In this process, however, they undergo spectacular changes, growing in radius and luminosity as giant stars, while casting off much of their mass which becomes briefly visible as a planetary nebula. These changes---large stellar radius, high luminosity, and mass loss---have significant effects on any planets, asteroids, comets, and other bodies orbiting the star, which are described in this chapter.

Understanding the changes wrought by giant stars on their planetary systems is important for a number of reasons. On a cultural level, we want to know what the future of our own planet will be, and that of any distant descendants we may leave behind. Scientifically, the study of giant stars offers a way to probe the population of planets around stars slightly more massive than the Sun, which are not readily amenable to radial-velocity planet surveys when they are on the main sequence; but the population of planets orbiting such giant stars has evolved somewhat from what it was on the main sequence. And perhaps most significantly, there is now a growing body of evidence for planetary systems orbiting white dwarf stars. This evidence comes in the form of:
\begin{enumerate}
    \item a handful of detections of planets and dusty structures such as disintegrating asteroids;
    \item many discs of dust and gas surrounding white dwarfs which are thought to be formed from the remnants of asteroids or other bodies that passed too close to the star;
    \item the very high fraction of white dwarfs that are observed to have rock-forming elements in their atmospheres, again thought to originate from such disrupted bodies.
\end{enumerate}
Study of these can tell us about the bulk abundances of extra-Solar asteroids; but we must always be mindful that bodies orbiting white dwarfs may have been altered by the significant changes that the stars undergo as giants.

This Chapter begins by summarising the relevant parts of stellar evolution theory (Section~\ref{sec:evolution}) and celestial mechanics (Section~\ref{sec:celmech}) needed to understand the effects on planetary systems of stellar RGB and AGB evolution. I then describe the strong tidal forces induced by stellar radius expansion (Section~\ref{sec:tides}), and how these compete with stellar mass loss which tends to cause orbits to expand (Section~\ref{sec:2bodymassloss}). After that, I discuss the effects of mass loss in destabilising systems, which can explain the source of planetary material later detected close to or accreted onto white dwarfs (Section~\ref{sec:nbodymassloss}). I finally discuss the effects the increase stellar irradiation has both on bodies' orbits and on their physical properties (Section~\ref{sec:radiative}).

\section{Overview of relevant stellar evolution}\label{sec:evolution}

\begin{figure}
    \centering
    \includegraphics[width=\textwidth]{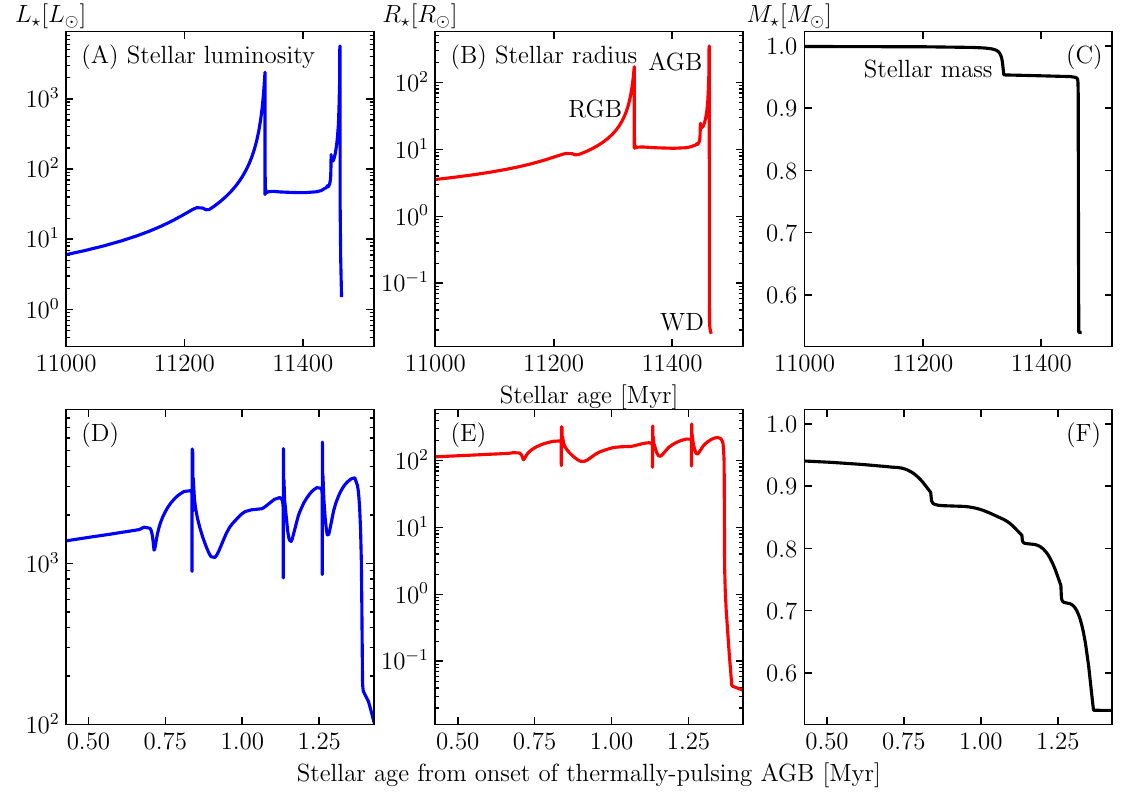}
    \caption{Evolution of a $1\mathrm{\,M}_\odot$ star with the same chemical composition as the Sun, from MIST stellar evolutionary tracks \citep{Choi16}. Top row: stellar luminosity (A), radius (B) and mass (C) for the final $\sim500$\,Myr before the star becomes a white dwarf, encompassing both the RGB and the AGB. Note the large increases in luminosity and radius during the AGB and the RGB, as well as the strong mass loss especially during the AGB. Bottom row: the same, for 1\,Myr around the AGB tip, showing the strong variations in luminosity (D) and radius (E), as well as the mass loss (F), during the thermal pulses.}
    \label{fig:evolution}
\end{figure}

We begin by revising some basic features of stellar evolution for low- and intermediate-mass stars. A more detailed treatment can be found in textbooks on stellar evolution, such as that by \cite{Prialnik09}. We shall be concerned exclusively with the progenitors of WDs, meaning stars of mass $M_\star\lesssim8\mathrm{\,M_\odot}$, and more particularly with the range $1-3\mathrm{\,M_\odot}$. The progenitor host stars of most evolved planetary systems would have lain in this mass range, whether we are considering planet-hosting giant stars, or WDs observed with photospheric pollution or other signs of remnant planetary systems. Stars more than $3\mathrm{\,M_\odot}$ are comparatively rare owing to the steepness of the stellar initial mass function, and some evidence suggests that planets may become less common around the more massive stars \citep{Reffert15}. At the lower mass end, few stars $<1\mathrm{\,M_\odot}$ have had the chance to evolve to become WDs within the age of the Universe, particularly when excluding metal-poor stars that have a lower likelihood of hosting planets. Finally, we also confine our attention to stars that are ``effectively single'' for the purposes of their evolution; i.e., they either have no stellar companion, or they have only a wide-orbit companion that does not affect the evolution of the star. Thus we exclude from consideration the future evolution of circumbinary planetary systems, which has received comparatively little attention.

A star spends most of its life as a \emph{main-sequence (MS)} object, fusing hydrogen into helium in its core, and most known planetary systems orbit such MS stars. The MS ends when hydrogen is exhausted in the core; at this point, the core collapses while hydrogen burning continues in a shell around the core. The combination of core collapse plus shell hydrogen burning drives the star up the giant branch, which can be partially understood as a consequence of the virial theorem \citep{Prialnik09}: as the core shrinks and heats up, the envelope expands and cools, while the luminosity rises significantly (see Figure~\ref{fig:evolution}A and B). The high luminosity, combined with the high opacity in the cool envelope, leads to the envelope becoming unstable to convective gas motions. We now refer to the star as a \emph{red giant branch (RGB)} star, and the large radius of the convective envelope plays a crucial role in strengthening tidal interactions, described later. 

Eventually, RGB evolution ends when helium ignites in the core. There is a qualitative difference in RGB evolution at around  $2\mathrm{\,M_\odot}$: above this limit, core helium burning begins to occur gradually in a non-degenerate core before the star has ascended far up the RGB. Below this limit, core temperature and pressure are too low for He burning to begin before the core becomes degenerate; when it starts, a runaway \emph{helium flash} occurs which abruptly terminates the RGB after a much larger radius ($\sim1$\,au) has been attained. After a relatively brief time, core helium is exhausted, and a second phase of core collapse and shell burning drives star back up the giant branch as an \emph{asymptotic giant branch (AGB)} star, attaining still larger radii and luminosities than on the RGB. Again, the envelopes of these objects are deeply convective. Towards the tip of the AGB, \emph{thermal pulses} occur where nuclear reactions in the hydrogen- and helium-burning shells alternatively switch on and off, leading to large oscillations at intervals of of $\sim10^5$\,yr in luminosity and radius imposed on the overall growth trend (Figure~\ref{fig:evolution}D and E).

The combination of the large stellar radius, the high luminosity, and the high opacity on the RGB and particularly the AGB means that the envelopes of these stars are rather weakly bound. While the Sun is currently losing mass at a rate of only $10^{-14}\mathrm{\,M_\odot\,yr}^{-1}$, mass loss rates around giant stars can reach $10^{-7}\mathrm{\,M_\odot\,yr}^{-1}$ at the RGB tip and $\sim10^{-4}\mathrm{\,M_\odot\,yr}^{-1}$ at the AGB tip in the \emph{superwind} phase. While these phases of strong winds are relatively brief, they dominate the overall mass loss over a star's life: the Sun will lose only $\sim0.1\%$ of its mass over its 10\,Gyr MS lifetime, but will overall lose around half its mass in becoming a WD, mostly at the AGB tip (Figure~\ref{fig:evolution}C and F). Modern stellar evolution models have much of the mass loss taking place towards the AGB tip \citep[e.g., MESA/MIST][]{Choi16}, but the reader should bear in mind that some literature on modelling planetary systems around evolving stars makes use of older stellar evolution models (such as SSE) which had more of a balance between the total mass lost on the RGB and the AGB for low-mass stars. Models of intermediate-mass stars which do not undergo a core helium flash always result in most mass loss occurring towards the AGB tip.

After the AGB tip and the final stages of mass loss, a young WD is left behind surrounded by the expelled envelope which is illuminated to become briefly visible as a \emph{planetary nebula}. 

We now provide some rough rules of thumb to aid the reader's intuition in quantifying these stellar evolutionary processes. A $1\mathrm{\,M_\odot}$ star has a MS lifetime of order $10^{10}$\,yr. The MS luminosity is a steep function of mass owing to the sensitive temperature dependence of nuclear fusion reactions; hence, the MS lifetime decreases significantly with increasing stellar mass, and a $2\mathrm{\,M_\odot}$ star will spend only $\sim10^9$\,yr on the MS. As an RGB star, a $1\mathrm{\,M_\odot}$ star will attain a radius of around 1\,au, with a $\gtrsim2\mathrm{\,M_\odot}$ star above the mass for nondegenerate He burning attaining a smaller radius. At the AGB tip, the maximum radius increases with stellar mass, with a $1\mathrm{\,M_\odot}$ AGB tip star attaining a radius of over 1\,au and a $3\mathrm{\,M_\odot}$ AGB tip star attaining a radius of around 3\,au. The fraction of mass lost also grows with stellar mass (although more massive stars still give rise to slightly more massive WDs), with a $1\mathrm{\,M_\odot}$ star losing roughly half its mass and a $3\mathrm{\,M_\odot}$ star losing roughly three-quarters. These values also depend somewhat on the stellar metallicity, but rather weakly over the metallicity range important for planetary systems.

The MS lifetime is a steep function of stellar mass. Many ``old'' or ``evolved'' planetary systems orbiting giants or WDs are therefore actually younger than the Solar System's 4.6\,Gyr! They can be younger in dynamical terms (planetary orbits completed) than the Solar System, while any potentially habitable planets may not have had a chance to develop complex life.

\section{Overview of orbital dynamics}

\label{sec:celmech}

\subsection{Two-body Keplerian orbits}

\label{sec:2body}

\begin{figure}
    \centering
    \includegraphics[width=\textwidth]{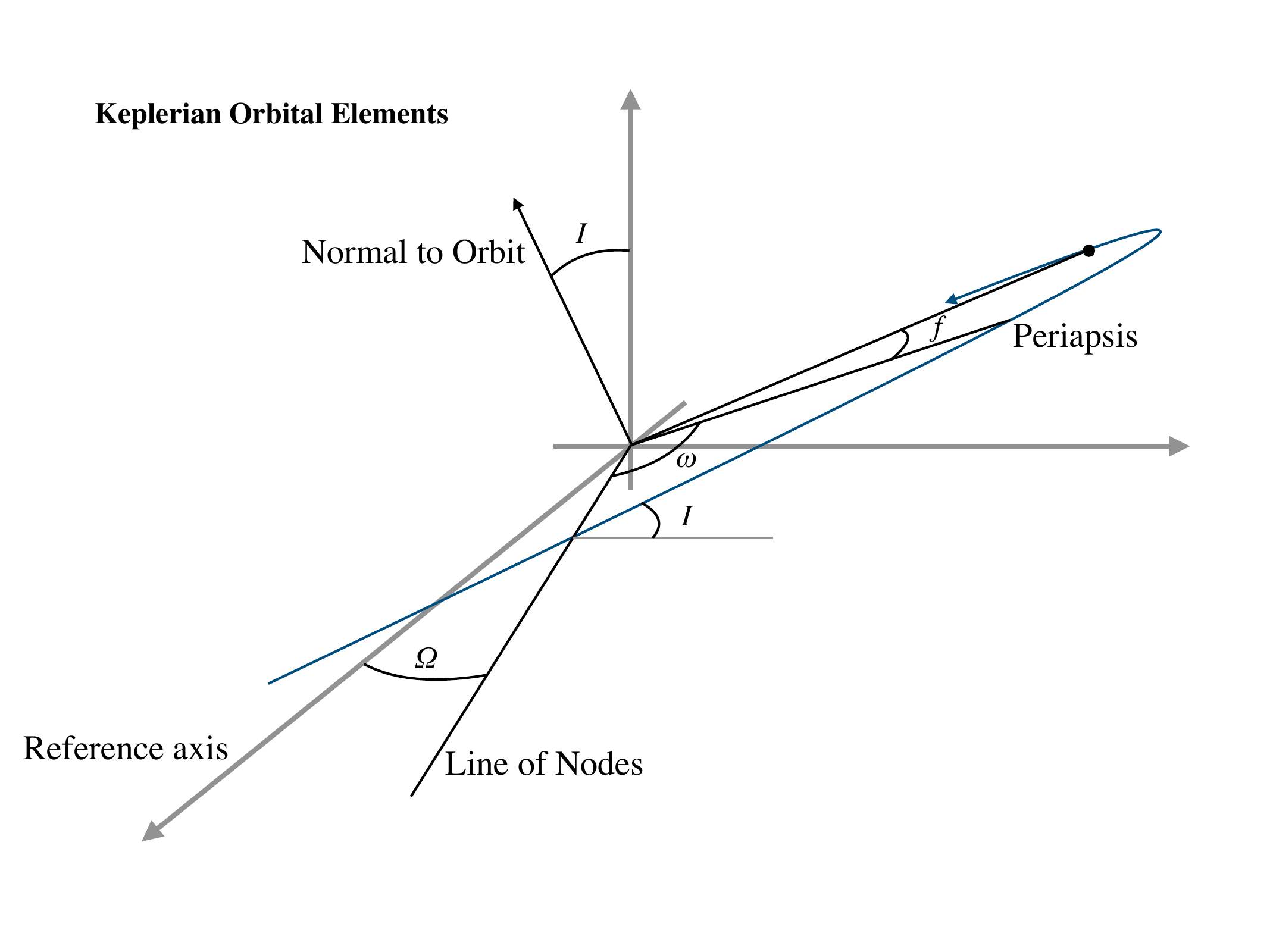}
    \caption{Keplerian orbital elements inclination $I$, longitude of ascending node $\Omega$, argument of periapsis $\omega$, and true anomaly $f$, for a bound elliptical orbit.}
    \label{fig:orbelts}
\end{figure}

It is a happy coincidence that the gravitational potential outside a sphere of matter is the same as that generated by a point mass of equivalent mass located at its barycentre: the \emph{Kepler potential}
\begin{equation}
    \Phi(r) = -\frac{\mathcal{G}M_\star}{r},
\end{equation} 
where $\mathcal{G}$ is the gravitational constant and $M_\star$ the stellar mass. Thus the motion of two spherical bodies is equivalent to the motion of a body orbiting a point mass. In this case, as in any static potential, a body's position and velocity at a given time uniquely determine its future orbit, an orbit thus defined by six independent parameters. While position and velocity may specify an orbit, they are constantly changing with time. It is therefore considerably more convenient to use the \emph{Keplerian orbital elements,} which relate more intuitively to the geometry of the orbit which, for a bound orbit, is elliptical. These orbital elements are:
\begin{itemize}
    \item the \emph{semimajor axis} $a$ (half the length of the ellipse);
    \item the \emph{eccentricity} $e$ (deviation from circularity of the orbit);
    \item the \emph{longitude or argument of periapsis}\footnote{Also called the \emph{periapse, pericentre,} or, for orbits around specific bodies, \emph{periastron, perihelion, perigee} etc.} $\varpi$ or $\omega$ (the orientation of the ellipse, relative to a fixed reference axis or to the intersection of the orbital and reference planes);
    \item the \emph{inclination} $I$ and \emph{longitude of ascending node} $\Omega$ (two angles that together specify the 3-dimensional orientation of the orbital plane);
    \item one angle to parametrise the orbital phase, common choices being the \emph{true anomaly} $f$ or \emph{true longitude} $\theta$ (the physical angle round the orbit, measured from the pericentre or a fixed reference axis respectively), or the \emph{mean anomaly} $\mathcal{M}$ or \emph{mean longitude} $\lambda$ (corresponding angles which increase linearly in time).
\end{itemize}
Several of these are illustrated in Figure~\ref{fig:orbelts}.

The utility of the Keplerian orbital elements is that all, save the mean anomaly (or equivalent parametrisation of orbital phase), are constant in the Kepler problem. This arises because of symmetries of the potential. We recall \emph{Noether's Theorem} from classical mechanics, which informally states that for every continuous symmetry of a system there is a corresponding conserved quantity. Conservation of energy arises from time invariance of the potential, and for the Kepler problem corresponds to a constant semimajor axis:
\begin{equation}
    E = -\frac{\mathcal{G}M_\star m_\mathrm{pl}}{2a}, \label{eq:Ekep}
\end{equation}
where $m_\mathrm{pl}$ is the planetary mass. Conservation of the angular momentum vector arises from spherical symmetry, and corresponds in the Kepler problem to fixed orbital eccentricity, inclination, and longitude of ascending node; the magnitude of the angular momentum vector can be expressed as
\begin{equation}
    L = m_\mathrm{pl}\sqrt{\mathcal{G}M_\star a\left(1-e^2\right)}. \label{eq:angmom}
\end{equation}
Any spherically-symmetric potential will place these four independent constraints on particles' orbits, confining each orbit to its own plane. A near-unique property of the Kepler potential is the existence of a further ``hidden'' mathematical symmetry, which gives rise to invariance of the longitude of periapsis. These symmetries then place five constraints on the Keplerian orbit, confining it to a one-dimensional manifold: as it happens, a conic section. The one degree of freedom permitted is then the orbital phase.

Evolution of these orbital elements arises from symmetry-breaking; one example highly relevant for this chapter is the change to the potential caused by stellar mass loss, which breaks time invariance and hence energy conservation. Additional effects that can break the symmetries of the Kepler potential and induce orbital evolution include relativistic effects, deviations from sphericity of one or both bodies (such as tidal deformation), and radiative forces; and a major class of such effects is the impact of additional massive bodies in the system such as planets or stars, and we deal with these in the following subsection. Usually these extra effects are weak compared to the dominant Keplerian motion: a body follows an approximately Keplerian orbit on an orbital timescale, and we study slow, long-term changes to the orbital elements. Sometimes, however, elements can change near-impulsively, such as when two planets experience a close encounter, or when mass loss is extremely rapid (as happens in binary systems where one component undergoes a supernova).

\subsection{Systems of $N>2$ bodies}

\label{sec:nbody}

The introduction of an additional massive body into a two-body system complicates the dynamics considerably, and general analytical solutions do not exist. For analytical work, we must make use of various approximations with different ranges of applicability. Alternatively, we may simulate orbital evolution numerically on a computer; here, the analytical approximations remain useful to aid interpretation and understanding. Textbook treatments of this area include those by \cite{MurrayDermott99}, covering analytical aspects, and more recently by \cite{Tremaine23}, who additionally covers numerical integrations.

Systems of multiple bodies may broadly be stable or unstable. By ``stable'' we usually mean simply that the system maintains the same qualitative behaviour for some long but not infinite time of interest (such as the lifetime of a star, or of the Universe), while we speak informally of a system ``becoming unstable'' or ``undergoing an instability'' when a significant change happens to a body or the system as a whole, such as the excitement of large eccentricities leading to orbits intersecting\footnote{This differs from the formal mathematical notion of instability as motion away from a fixed point or limit cycle: all informally unstable systems are mathematically unstable, but so also may be many informally stable systems.}. The outcome of orbital instability may be the ejection of one or more bodies on an unbound orbit, or the close approach of two bodies, which itself may lead to a physical collision, orbital change through strong tidal interactions, or the disruption of a body by a strong tidal force.

\subsubsection{Strong interactions: close encounters and scattering}

A simple but important case of three-body dynamics occurs when two bodies undergo a \emph{close encounter} or gravitational \emph{scattering} event, often as a result of the other dynamical processes to be described below. For example, two planets can scatter each other, or a planet can scatter an asteroid or comet. The region around a planet where this process is significant can be described by the \emph{Hill sphere,} whose radius (the \emph{Hill radius}) approximates the volume where the planetary gravitational potential dominates over the stellar and centrifugal potentials. The Hill radius is given by
\begin{equation}
    r_\mathrm{H} = a\left(\frac{m_\mathrm{pl}}{3M_\star}\right)^{1/3}. \label{eq:rH}
\end{equation}
Roughly speaking, outside the Hill sphere a body's orbital motion is dominated by the Keplerian potential of the star, while inside the Hill sphere it is dominated by the Keplerian potential of the planet. An orbit inside the Hill sphere may be a bound circle or ellipse, as in the case of a satellite, or a hyperbola in the case of a scattering event\footnote{A more sophisticated treatment, \emph{Hill's approximation,} also takes into account the Keplerian motion of both bodies around the star.}. The outcome of such a scattering event depends on the relative masses and velocities of all bodies, as well as the impact parameter of the event, but broadly the outcomes are either a deflection of the orbit, or collision of the two bodies\footnote{Readers coming from other sub-fields, such as the dynamics of stellar clusters or galaxies, should note that in planetary dynamics the term ``collision'' refers specifically to a physical impact, rather than to any gravitational scattering as in a ``collisional'' stellar system.}; tidal disruption of the less dense body may also be possible. A planet has a larger cross section for scattering versus collision the larger its Safronov number is, where the Safronov number is given by 
\begin{equation}
    \Theta = \frac{2m_\mathrm{pl}}{M_\mathrm{\star}}\frac{a}{R_\mathrm{pl}} = \left(\frac{v_\mathrm{esc}}{v_\mathrm{Kep}}\right)^2 \label{eq:safronov}
\end{equation}
which is the square of the ratio of the planet's surface escape velocity to its Keplerian orbital velocity \citep{Tremaine23}. In Equation~\ref{eq:safronov}, $R_\mathrm{pl}$ is the planet's physical radius. A higher $\Theta$ means a planet is more prone to scattering as a result of a close encounter, while a small $\Theta$ means it is more prone to collision. A more massive planet can induce a larger velocity change during an encounter without experiencing a physical collision. To provide some examples, a low-mass planet on a relatively close-in orbit, such as Earth, has a small Safronov number ($v_\mathrm{Kep}=30\mathrm{\,km\,s}^{-1}$, $v_\mathrm{esc}=11\mathrm{\,km\,s}^{-1}$) and is relatively more efficient at accreting bodies, and less efficient at scattering them, than a more massive planet on a wider orbit such as Jupiter ($v_\mathrm{Kep}=13\mathrm{\,km\,s}^{-1}$, $v_\mathrm{esc}=60\mathrm{\,km\,s}^{-1}$).

The single-planet Hill radius defined above is suitable for use when one body is of negligible mass; for two planets of comparable mass, we can define similarly the \emph{mutual Hill radius}
\begin{equation}
r_\mathrm{H,mut} = \frac{a_1+a_2}{2}\left(\frac{m_1+m_2}{3M_\star}\right)^{1/3}. \label{eq:rhmut}
\end{equation}
Using the mutual Hill radius, there is a particularly simple expression for the stability of two planets initially on circular orbits: their orbits may not intersect if
\begin{equation}
    \Delta > 3.5R_\mathrm{H} \label{eq:gladman}
\end{equation}
\citep{Gladman93}. Note that this criterion for \emph{Hill stability} does not preclude that one planet escape to infinity, or collide with the star.

\subsubsection{Long-term evolution}

In many systems, gravitational scattering forms only a part of the dynamics, or may not occur at all. Most of the evolution occurs under weaker, slower-acting interactions, with orbital elements changing gradually, which may subsequently lead to the close encounters described above. In such cases, we often make use of \emph{osculating} orbital elements, which are those instantaneously derived from a body's position and velocity, and which may be taken as fixed on an orbital timescale. To calculate the changes on long timescales, we are uninterested in short-period fluctuations owing to planetary conjunctions, which typically average out in the long run, and so we seek expressions for the averaged perturbing potential a body feels from another planet or star, as a function of orbital elements rather than Cartesian position. This perturbing potential can be expressed in several ways. One approach, common for near-coplanar systems such as the Solar System, is to utilise Fourier series where the angles are combinations of orbital longitudes (mean, apsidal, and nodal) and the coefficients are functions of semimajor axes, eccentricities and inclinations. Here the potential takes the form \citep[see, for example,][Chapter~6]{MurrayDermott99}
\begin{equation}
    R = \sum_\phi S(a_1,a_2,e_1,e_2,I_1,I_2) \cos \phi,
\end{equation}
for coefficients $S$ and angles $\phi$ where
\begin{equation}
    \phi = j_1\lambda_1 + j_2\lambda_2 + j_3\varpi_1 + j_4\varpi_2 + j_5\Omega_1 + j_6\Omega_2,
\end{equation}
where $j_i$ are integers summing to zero\footnote{This ensures invariance of the angle $\phi$ to the choice of reference axis.}. The coefficients $S$ are  expressible as power series in the small parameters of eccentricity and inclination, and this is therefore unsuitable for high inclinations and eccentricities. As a suitable alternative for this situation, a multipole expansion may be performed, valid for arbitrary eccentricity and inclination but now with the semimajor axis ratio as a small parameter; this is useful for scenarios such as planets in binary systems experiencing the Kozai effect. These series are averaged over the mean longitudes, and only significant terms retained in order to study long-term evolution.

Under orbit-averaging, most of the Fourier terms involving mean longitudes vanish, as they vary rapidly over the course of an orbit. The exception comes if two bodies' periods are close to an integer commensurability, such as 2:1. In this case, the terms involving the angles $\phi_1=\lambda_1-2\lambda_2+\varpi_1$ and $\phi_2=\lambda_1-2\lambda_2+\varpi_2$ would not average out on an orbital timescale, as the two angles are slowly varying. Such a situation is known as a \emph{mean motion resonance (MMR)}. In this configuration, conjunctions between bodies occur close to the same locations in inertial space, which means configurations of close or even intersecting orbits can remain stable, as in the case of Neptune and Pluto. This also means that the kicks bodies receive from each other at conjunction add coherently, leading to significant dynamical interactions on relatively short timescales: the semimajor axis, eccentricity and resonant angle typically librate around a fixed point. On long timescales, mean motion resonances can be either stable or unstable, depending on the details of the resonance and how it interacts with the global properties of the planetary system: witness the stable resonances of Trans-Neptunian Objects such as Pluto with Neptune, versus the unstable \emph{Kirkwood gaps} sculpted by Jupiter's mean motion resonances in the asteroid belt.  MMRs are stronger the smaller the difference between integers, so first order MMRs such as 2:1, 3:2 etc are usually more significant than high-order resonances such as 11:4. MMRs have a finite width in semimajor axis, which depends on eccentricity and on the planet:star mass ratio
\begin{equation}
\Delta a_\mathrm{res} \propto \left(\frac{m_\mathrm{pl}}{M_\star}\right)^{1/2} \label{eq:res}
\end{equation}
for moderate eccentricities. Instability of MMRs is linked to the overlap of resonances, and/or to secondary resonances with other dynamical frequencies in the planetary system \citep{Lecar01}. The overlap of neighbouring mean-motion resonances leads to chaotic behaviour (usually implying orbital instability) in a \emph{chaotic zone} of width
\begin{equation}
    \Delta a_\mathrm{chaos} = 1.4\left(\frac{m_\mathrm{pl}}{M_\star}\right)^{2/7} \label{eq:chaos}
\end{equation}
surrounding the planet's orbit.

The interplay of different mean motion resonances determines the stability of systems of multiple planets. When there are three or more planets in a system, additional three-planet resonances play a role, where the resonant angles are given by $\phi=j_1\lambda_1+j_2\lambda_2+j_3\lambda_3$; a familiar example from our Solar System is the 4:2:1 \emph{Laplace resonance} between Io, Europa and Ganymede. These three-planet resonances form a ``web'' in the parameter space of bodies' semimajor axes, through which the system slowly diffuses in chaotic motion \citep{Petit20}. Instability can then occur when the system reaches a stronger two-planet MMR. The typical lifetime for this to happen is given by
\begin{equation}
    t \propto \left(\frac{m_\mathrm{pl}}{M_\star}\right)^{1/4}, \label{eq:3planet}
\end{equation}
which has a slightly weaker mass scaling than the mutual Hill radius often historically used to characterise the dynamical spacing of multi-planet systems.

Finally, we turn to perturbations independent of bodies' mean longitudes, dependent only on the longitudes of periapsis and node. These, known as \emph{secular} perturbations, are always present to some level in any multi-planet system. To lowest order, a secular system can be regarded as a system of coupled harmonic oscillators where the planets' eccentricities and inclinations are oscillating. The eigenfrequencies of this system play a significant role in its dynamics: they set the frequencies for oscillation of the planets' eccentricities and inclinations, as well as the corresponding precession rates of their apsides and nodes. Strong instability can occur when one of these eigenfrequencies matches another frequency in the system. This occurs at \emph{secular resonances,} where the precession frequency of one body forced by the system's secular perturbations matches one of the secular eigenfrequencies. It also occurs at \emph{secondary resonances,} where a secular precession frequency matches the libration frequency of a MMR \citep{Lecar01}. Secular and secondary resonances are responsible for carving the structure of the asteroid belt, its edges and Kirkwood gaps.

In systems of very high eccentricity or inclination, a multipole expansion of the perturbing potential is appropriate. In such a expansion, the leading-order term is a quadrupole\footnote{Unlike the case of electrostatics, there can be no negative mass to give rise to a gravitational dipole.}. The effects of this quadrupolar potential are known as the \emph{Kozai effect}\footnote{Having been (re-)discovered by several authors, it also also known as the Lidov--Kozai, or von Zeipel--Lidov--Kozai, effect.}. The Kozai effect leads to a change of the inner body's eccentricity and inclination while conserving the $z-$component of the orbital angular momentum. Starting from a circular orbit, high eccentricities can be attained if the initial mutual inclination is greater than $\approx39^\circ$. Inclusion of octupolar terms can lead to the excitation of yet higher eccentricities \citep{Naoz16}. This effect can be significant in planetary systems with wide binary companions, as well as planetary systems that attain large mutual inclinations during or after a phase of scattering.

Owing to the complexity of these effects---which often operate in series or in parallel within the same system---the planetary $N$-body problem is often studied numerically. The planetary $N-$body problem, compared to galactic or cosmological simulations, involves comparatively few bodies but often a huge duration in terms of dynamical timescales, which has motivated the development of specialised integrators and software packages. Common integration algorithms include \emph{symplectic integrators} (constructed so as to guarantee conservation of a quantity close to the system's true energy), and high-accuracy non-symplectic integrators such as RADAU (often in practice more accurate than the symplectic integrators). The effects of stellar evolution have been incorporated into commonly-used $N-$body packages such as \textsc{Mercury} \citep{Veras+13,Mustill+18} and \textsc{Reboundx} \citep{Baronett+22}. In so doing, to ensure good convergence, care must be taken to ensure that the changing stellar parameters are interpolated correctly within the integration timesteps, rather than merely being periodically updated \citep{Veras+13,Baronett+22}.

\section{Tidal orbital decay} 

\label{sec:tides}

Hitherto we have dealt with point or spherical masses generating and moving within each other's Keplerian potentials. However, any extended body in a gravitational potential feels a differential gravitational force across its volume; this is known as the tidal force. Here we consider the case of an extended giant star distorted by an orbiting planet, but the same physics applies to such systems as the planet distorted by the star, planets and their satellites, and stellar clusters orbiting the Galaxy. The tidal mechanism relevant for giant stars is the \emph{equilibrium tide,} where the star is distorted to align with equipotential contours generated by the combined gravity of the planet and the star in the reference frame rotating with the planet's orbital mean motion. An alternative class of tidal interaction, the \emph{dynamical tide,} involves the excitation and dissipation of oscillation modes in one or both bodies; this is not significant for giant stars.

\begin{figure}
    \centering
    \includegraphics[width=\textwidth]{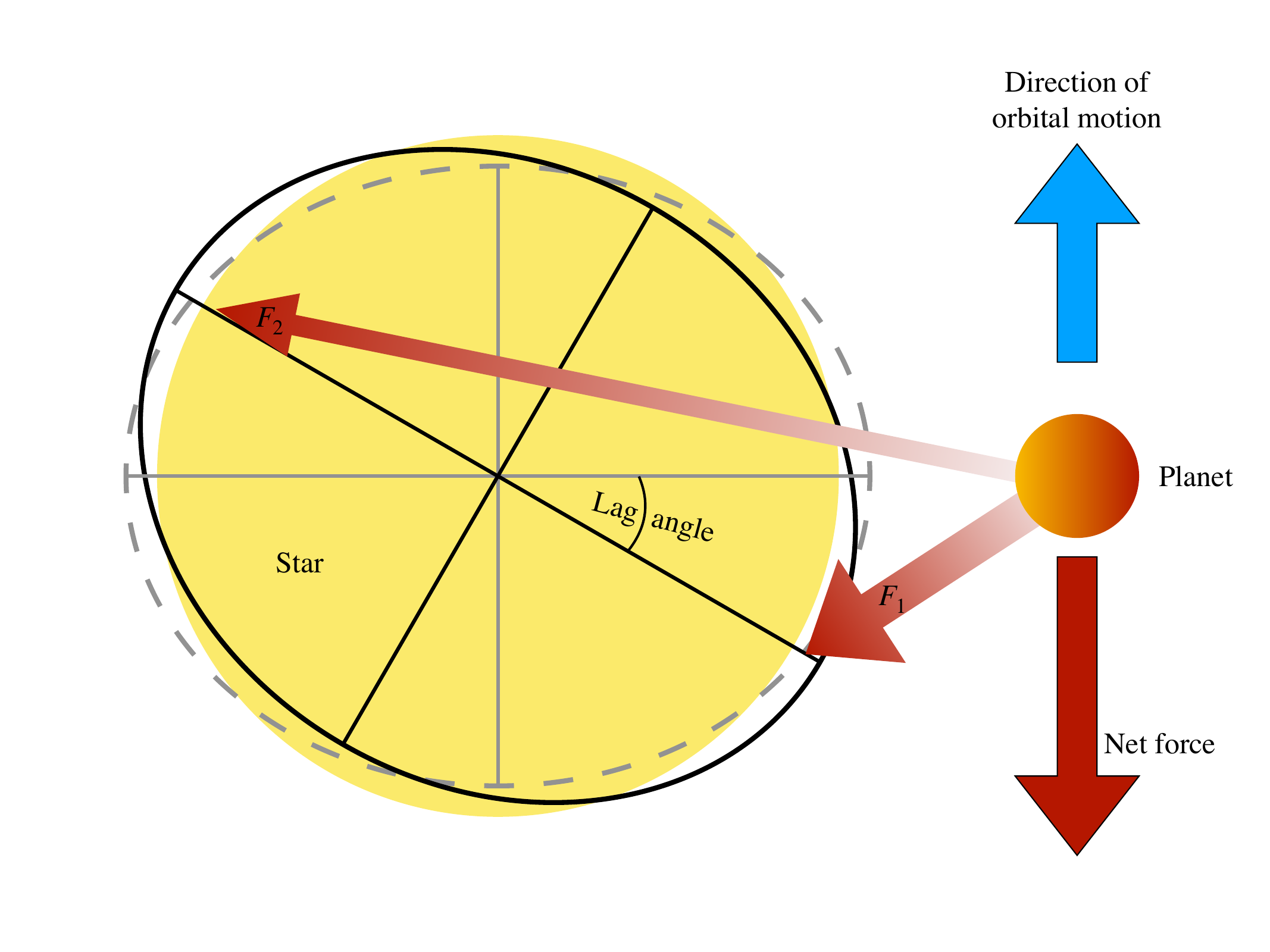}
    \caption{Cartoon illustrating the principle of the equilibrium tide. In a reference frame corotating with the planet's orbital motion, the undisturbed star (yellow) is distorted and attempts to fill an equipotential (grey dashed ellipse). As the star here is rotating more slowly than the planet orbits, the tidal bulge lags the planet's position, and the star instead aligns with the black ellipse (the lag angle here is greatly exaggerated for clarity). The planet then feels a stronger force from the nearer bulge ($F_1$) than from the further bulge ($F_2$), resulting in a net force opposing orbital motion. The outcome of this is the loss of orbital angular momentum and the inspiral of the planet towards the star.}
    \label{fig:tidal}
\end{figure}

The equilibrium tide mechanism is illustrated in Fig~\ref{fig:tidal}. The star being a fluid body, its surface attempts to fill an equipotential contour in the rotating reference frame, giving rise to a deformation that can be described as two bulges: one pointing towards the planet, and a second pointing away from the planet at the antipodal point. In general, the fluid of the stellar envelope has a finite response time to the changing potential, known as the \emph{lag time}. This means that the tidal bulges will not align exactly with the axis directed towards the planet, unless the stellar rotation period matches the planetary orbital period \emph{(tidal synchronisation).} Instead, the tidal bulges lag or lead this axis, depending on whether the stellar rotation is slower or faster than the planet's orbital motion. The lag time or angle depends on the nature of the energy dissipation in the star's envelope and in general may depend (often non-trivially) on the frequency of the forcing from the orbital motion. This misalignment of the tidal bulges results in a tangential force (and therefore torque) between the planet and the deformed star, which changes both the spin of the star and the orbit of the planet.

Giant stars have large radii and their envelopes have large moments of inertia, and so they typically spin very slowly. Therefore for planets orbiting giant stars that are subject to significant tidal forces, the planetary orbit is more rapid than the stellar spin and the tidal bulge lags the planet's orbital location. This removes angular momentum from the planet's orbit, causing the planet to slowly spiral inwards towards the star until it is eventually engulfed in the envelope. 

The exact mathematical derivation of the tidal torque is complex, but it is easy to appreciate that the torque must be an extremely strong function of the ratio of the star's radius $R$ to the planet's orbital semimajor axis $a$. The star feels a tidal potential which is a quadrupole, proportional to $(R/a)^3$, and it attempts to fill these equipotential surfaces. This small quadrupolar distortion of the star then generates its own potential that differs from the Kepler potential by a further $(r/R)^3$ quadrupole at a radius $r$ exterior to the star; at the orbital radius of the planet, these combine to give a perturbing potential felt by the planet proportional to $(R/a)^6$. The exact scaling with radius depends on the details of the dissipation mechanism; for giant stars, the energy of the tidal distortion is dissipated by convective motions in the stellar envelope, and the resulting rate of orbital decay is given by
\begin{equation}
    \left(\frac{\dot{a}}{a}\right) = \frac{f}{\tau}\frac{M_\mathrm{env}}{M_\star}q(1+q)\left(\frac{R_\star}{a}\right)^8,
\end{equation}
where $q=m_\mathrm{pl}/M_\star$, $\tau$ the convective turnover timescale, and $f$ a frequency-dependent term accounting for the efficiency with which the tidal motions couple to the convective motions. \citep{Zahn89,Villaver07}. Owing to the extremely strong dependence of the tidal torque on $R/a$, as $R_\star$ grows from the MS to the RGB or AGB tip by a factor of 100, tides go from utterly insignificant to of overriding importance for planets on orbits out to several au. The inwards orbital decay induced by tidal forces competes with orbital expansion induced by stellar mass loss, as discussed in the next section.

It is common to parametrise the strength of tidal dissipation within a body with a parameter $Q$, the \emph{tidal quality factor,} quantifying the relative amount of the energy stored in the distortion that is dissipated in a single cycle. A constant $Q$ is rarely a suitable model for long-term evolution, as the quality factor depends on forcing frequency often in a very non-trivial manner. However, an effective instantaneous $Q$ can be calculated and is instructive to compare the relative strengths of tides in different systems. For planets orbiting giant branch stars just prior to engulfment of a planet, the effective $Q$ is $\sim10^2$ \citep{Nordhaus+10}, meaning the strength of the tide in this situation is orders of magnitude stronger than the strength of a tide raised by a planet on a MS star ($Q\gtrsim10^7$, e.g., \citealt{CollierCameron18}).

\section{Stellar mass loss and its effects on orbital dynamics}

\subsection{Mass loss in two-body systems: orbit expansion and competition with tidal inspiral}

\label{sec:2bodymassloss}

We now consider the effects of stellar mass loss on orbiting bodies, beginning with the star--planet system covered comprehensively by \cite{Veras+11}. In most cases stellar mass loss does not create a significant perturbative force (though there are weak aerodynamic drag forces on very small bodies); instead, mass loss from the star causes a change to its gravitational potential, which then changes bodies' orbital elements. Recalling Noether's Theorem from Section~\ref{sec:2body}, we remember that conservation of energy in a system arises from a time symmetry. By allowing mass loss from the star, we now have a time-dependent potential, meaning that bodies' orbital energy is no longer conserved: in general, orbits expand as the star loses mass. Some constraints can, however, still be placed on the orbital evolution. Mass loss from RGB and AGB stars is unlikely to be strongly directional, meaning that rotational invariance of the potential is not broken and therefore that orbital angular momentum is conserved as the orbit expands\footnote{Although anisotropic effects may matter for bodies on very wide orbits, such as Oort cloud comets close to apoapsis.}. Isotropic mass loss also means that the orbital inclination and longitude of ascending node do not change. In the important case of \emph{adiabatic mass loss,} where the time-scale for mass loss is significantly longer than the orbital timescale, the orbital eccentricity is also an \emph{adiabatic invariant} and does not change as mass loss proceeds; nor does the argument of periapsis. The orbit therefore expands while maintaining the same shape, and from conservation of angular momentum (Equation~\ref{eq:angmom}) we have the following relation between the initial and final semimajor axes:
\begin{equation}
    \frac{a_\mathrm{final}}{a_\mathrm{inital}} = \frac{M_{\star\mathrm{,initial}}}{M_{\star\mathrm{,final}}}.\label{eq:adiabatic}
\end{equation}
Thus, planets orbiting a $1\mathrm{\,M}_\odot$ star, which loses around half of its mass, will roughly double their orbital semimajor axes; those orbiting a $3\mathrm{\,M}_\odot$ star, which loses around three-quarters of its mass, will roughly quadruple theirs.

While mass loss is typically slow and adiabatic for planets on orbits of a few to tens of au, it can be significantly non-adiabatic for bodies with wider orbits and longer orbital periods, such as wide binary stellar companions and Oort cloud comets. Under non-adiabatic mass loss, orbital angular momentum is still conserved, but the eccentricity can change; the orbital semimajor axis and periapsis distance still increase, typically faster than in the adiabatic case \citep{Veras+11}. For sufficiently rapid mass loss, bodies can become unbound: such is for example the case for \emph{impulsive} (effectively instantaneous) mass loss such as that resulting from a supernova explosion, where the loss of over half of the star's mass will cause a body initially on a circular orbit to possess positive energy and hence to be on a hyperbolic orbit after the explosion. For the low- and intermediate-mass stars with which we are concerned, non-adiabatic mass loss leads to the unbinding of their Oort Cloud comets and their release into the Galaxy as interstellar objects (ISOs) such as 1I/`Oumuamua and 2I/Borisov.

\subsubsection{Mass loss vs tides}

\begin{figure}
    \centering
    \includegraphics{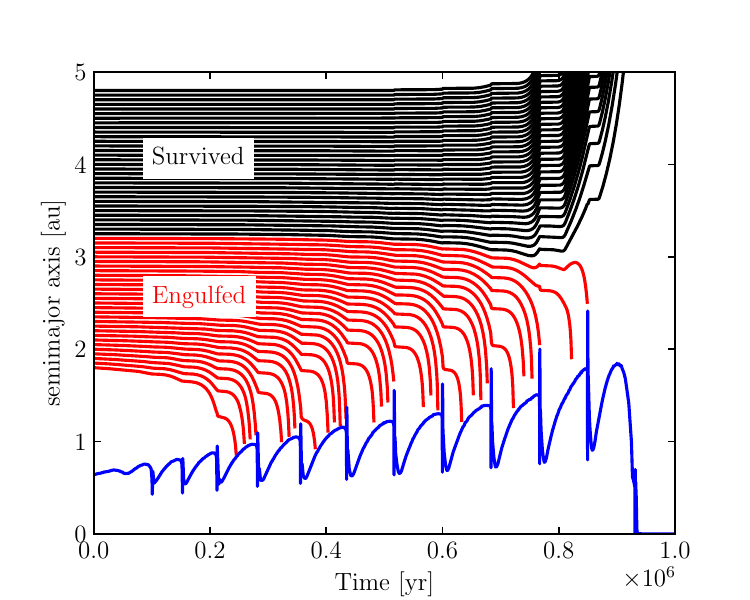}
    \caption{Orbital evolution of $1\mathrm{\,M_J}$ planets orbiting $1.5\mathrm{\,M}_\odot$ stars. A suite of one-planet systems is evolved through the thermally-pulsing AGB, each system having its planet placed on a circular orbit at a different initial semimajor axis. The planets' orbits evolve under the combined effects of stellar mass loss and tidal orbital decay, and are removed if they enter the stellar envelope (blue line). Planets beginning interior to $\sim3.2$\,au are engulfed (red lines), while those on wider orbits survive (black lines). Tidal models from \cite{Mustill12} with stellar evolution models from \cite{Vassiliadis93}.}
    \label{fig:agbtides}
\end{figure}

We see now that tidal forces and mass loss are in competition for planets initially orbiting at a few au: tides drag planets in towards the star, while mass loss causes their orbits to spiral outwards (Figure~\ref{fig:agbtides}). The balance between engulfment and survival has been probed in many studies under different assumptions about stellar evolution, mass loss and the strength of tidal forces \citep[e.g.,][]{Villaver07,Nordhaus+10,Mustill12,Sanderson+22}. Restricting attention to planets on circular orbits, when considering a planet of a given mass, the tidal force is a monotonically decreasing function of initial orbital radius; therefore, within a certain radius planets of that mass will be engulfed, while exterior to it they will survive. This critical radius depends on the mass of both the star and the planet, and is in the region of a few au \citep{Mustill12,Nordhaus13}. As the tidal forces decay very rapidly with orbital semimajor axis, most planets beyond the destruction radius will simply expand their orbits according to Equation~\ref{eq:adiabatic}. Over a relatively narrow range of intermediate separations, planets' orbits expand less than anticipated from mass loss alone (lowest few black lines in Figure~\ref{fig:agbtides}), or may even in rare cases shrink. However, most young WDs should be surrounded by a cleared volume of at least several au in radius, where no planets are found. The extent of this volume depends strongly on stellar progenitor mass: it is larger for the more massive WDs with more massive progenitors, which attain greater AGB radii and lose a larger fraction of their mass. It also depends on the modelled strength of stellar mass loss and tidal effects. In principle these models could be tested with a large population of planets detected at suitable distances (\emph{i.e.,} a few to ten au) around WDs. While JWST has recently imaged some giant planet candidates around WDs, the best way to detect large numbers of these planets may be astrometrically in future data releases of the \emph{Gaia} spacecraft. However, although \emph{Gaia} is expected to yield detections of planets around WDs they will likely not be numerous enough to meaningfully constrain the tidal models \citep{Sanderson+22}.

Once engulfed, planets' orbits shrink rapidly owing to gravitational friction with the stellar envelope, taking $<100$\,yr to reach $1\mathrm{\,R}_\odot$ \citep{OConnor23}: this is essentially a form of \emph{common-envelope (CE) evolution} with an extremely low-mass companion. While a stellar companion can survive such an event, transferring its orbital energy to the stellar envelope and expelling it, and ending up as a short-period \emph{post-common envelope binary,} the prospects for planets are much bleaker. They lack enough energy to eject the stellar envelope, meaning that their inspiral is not terminated by ejection of the envelope and continues until the planet is destroyed. This happens either by evaporation within the denser regions of the stellar envelope, or by tidal disruption at the Roche limit at $\sim1\mathrm{\,R}_\odot$. The lower mass limit for survival of bodies inside the stellar envelope is thought to be of the order of $10\mathrm{\,M_J}$ depending on factors such as the efficiency of energy transfer in the CE phase, and is coincidentally around the planet/brown dwarf mass boundary \citep{Villaver07,Nordhaus13,OConnor23}.

The fact that planets and other material are in fact seen on tight orbits close to WDs shows that many systems undergo significant dynamical evolution throughout the WD's lifetime. This can be triggered by the dynamical changes in multi-planet systems during the AGB, as discussed next.

\subsection{Mass loss in multi-body systems: changes to dynamics and stability}

\label{sec:nbodymassloss}

Planets in systems of multiple planets, as those in single-planet systems, experience adiabatic or non-adiabatic orbit expansion in accordance with their semimajor axes. In systems spanning a large range of semimajor axes, it is possible for closer bodies to be in the adiabatic regime and more distant ones in the non-adiabatic regime, leading to an increase in semimajor axis ratios. For planetary systems located within a few tens of au, however, orbit expansion will be adiabatic for all bodies and the system will expand in a self-similar manner.

In multi-planet systems, the planet--star mass ratio increases following mass loss, by a factor of 2--4 for 1--3\,$\mathrm{M}_\odot$ progenitor stars. This has significant effects on the system's dynamics, since most of the dynamical interactions described in Section~\ref{sec:nbody} have a strength that depends on some power of this ratio. Speaking generally, following mass loss the planets are heavier relative to the star than they were before; this strengthens planet--planet and planet--asteroid interactions and leads to the destabilisation of bodies, or indeed entire systems, that survived their MS lifetime \citep{Debes02}. This is now thought to be responsible for the delivery of material close to the WD that is observed as photometric transits, dust and gas discs, and ultimately photospheric pollution. In this paradigm, asteroids, planets, or moons are destabilised from their original orbits and have their eccentricities excited to near unity. Their orbital periapsis then lies interior to the \emph{Roche limit} or \emph{tidal disruption limit} 
\begin{equation}
    R_\mathrm{Roche} = \left(\frac{3\rho_\star}{\rho_\mathrm{pl}}\right)^{1/3}R_\star
\end{equation}
around the WD, where $\rho_\star$ and $\rho_\mathrm{pl}$ are the average stellar and planetary densities, resulting in their disruption as tidal forces overcome internal binding forces \citep{MurrayDermott99}.

There are now numerous studies of the details of how different dynamical mechanisms are affected by stellar mass loss, of which there is only space to mention a few here. In systems of a single planet and a disc of comets/asteroids, unstable MMRs broaden during stellar mass loss [width $\propto\left(m_\mathrm{pl}/M_\star\right)^{1/2}$; Equation~\ref{eq:res}], and MMRs interior to a planet's orbit can be an efficient deliverer of material to the WD, although the extent of a disc affected is rather small \citep{Debes12}. Where the planet is close enough to the star to be affected by tidal forces, MMRs migrate along with the planet, sweeping through the system and resulting in more complex dynamics that can enhance or reduce the survivability of interior bodies \citep{Ronco20}. The planet's chaotic zone, with its width scaling as $\left(m_\mathrm{pl}/M_\star\right)^{2/7}$ (Equation~\ref{eq:chaos}), also broadens with mass loss \citep{Bonsor11}, although an external disc configuration (such as Neptune and the Kuiper Belt) is inefficient at delivering material close to the star and requires the presence of extra planets to deliver destabilised bodies to the WD \citep{Bonsor11}. 

In systems of multiple planets, the stability scalings in two-planet systems [scaling as the Hill radius $\propto\left(m_\mathrm{pl}/M_\star\right)^{1/3}$; Equation~\ref{eq:rhmut}, \ref{eq:gladman}] and in three-planet systems [scaling with $\left(m_\mathrm{pl}/M_\star\right)^{1/4}$; Equation~\ref{eq:3planet}] are affected by mass loss, leading to the destabilisation of entire systems that survived MS evolution \citep[e.g.,][]{Debes02,Veras+13,Mustill+18,Maldonado21}. In terms of both the efficiency and longevity of the delivery of material to a WD, destabilisation of multi-planet systems can be more effective than the moderate growth of chaotic regions in single-planet systems. For planets on wide orbits around WDs, the Safronov number (Equation~\ref{eq:safronov}) is high, favouring strong gravitational scattering and excitation of planetary eccentricities over collisions between planets once instability has occurred. The high-eccentricity planets can then interact with and destabilise entire planetesimal discs, rather than the relatively narrow regions affected by the growth of chaotic zones and Kirkwood gaps \citep{Mustill+18}. Instability in systems of giant planets resolves rather quickly, and most bodies are ejected rather than being forced down to the WD. However, in systems of low-mass (e.g., Earth- or Neptune-mass) planets the planets are inefficient at ejecting either one another or asteroids. Instead, the gravitational scattering can continue for many Gyr as the planetesimal disc is slowly depleted through a combination of ejection from the system and scattering down to the WD's Roche limit. Despite the relatively weak scaling of the stability timescale with planet--star mass ratio, high-multiplicity systems of low-mass planets may be sufficiently closely spaced that the majority of them can be destabilised by mass loss \citep{Maldonado21}, and they are therefore a good candidate to explain WD planetary phenomena on a population level.

Slightly different is the case of secular dynamics in planetary systems, where adiabatic mass loss \emph{per se} doesn't change the locations or widths of secular resonances (the changing planet--star mass ratio manifests instead in globally rescaling the secular frequencies). However, tidal decay or engulfment of the inner planets does change the location of secular resonances \citep{Mustill12,Smallwood21}, which can be highly efficient at exciting the near-unity eccentricities needed for delivery of bodies to the WD. A changed system architecture due to tidal migration can therefore place secular resonances in previously stable and well-populated regions of a planetesimal disc. Furthermore, radiative forces on the AGB can redistribute minor bodies through a planetary system (see Section~\ref{sec:radiative}), repopulating resonances that were destabilised on the MS. Finally, mass loss and tidal engulfment can change the nature of Lidov--Kozai interactions, for example by removing a planet interior to a disc that was suppressing Lidov--Kozai oscillations. Similar to the case of scattering by low-mass planets, such secular mechanisms can provide efficient delivery of asteroids to the WD over a long timescale \citep{Petrovich17}.

A major challenge in modelling these processes is our significant lack of knowledge about the relevant regions of planetary systems that survive RGB and AGB evolution, beyond a few au around stars more massive than the Sun. This state of ignorance applies to both the WD systems themselves and their MS and RGB/AGB progenitors. Our knowledge here is limited to a reasonable census of Jupiter-analogue planets on orbits of a few au, a handful of wide-orbit young massive planets detected through direct imaging, and a growing number of planet candidates detected indirectly from structures observed in protoplanetary discs, together with relatively good statistics on the broad occurrence rates of minor body populations in the form of debris disc studies around MS stars. The prevalence and multiplicity of lower-mass planets (Earth-mass, super-Earths and Neptunes), known to be extremely common on close-in orbits, is entirely unconstrained. In light of the weak observational data, theoretical studies can take three approaches to creating planetary systems on wider orbits. Most studies make synthetic systems based on simple prescriptions such as setting up 3 planets separated by a number of Hill radii \citep[e.g.,][]{Debes02,Veras+13,Mustill+18}. This has the advantage of control over system parameters, but the systems may not be representative of the real planetary population. \cite{Maldonado21} took a different approach, taking the well-studied systems close to the star (many being multi-planet systems discovered by the \emph{Kepler} spacecraft) and using them as templates, scaling them up to wider orbits. This implicitly assumes that processes of planet formation and migration are self-similar at sub-au distances around Solar-mass stars and at $\sim10$\,au around intermediate-mass stars. Finally, in principle one could consistently model the formation of a planetary system, including its minor body populations, and then study its long-term evolution under mass loss. However, planet formation models are not well constrained for planets on wide orbits around stars more massive than the Sun, and this has not been explored in practice.

\section{Radiative effects}

\label{sec:radiative}

\subsection{Radiative forces and their effects on orbits}

The motion of most bodies around stars is dominated by gravitational forces, both from the star and from other bodies in the system. However, \emph{radiative} or \emph{radiation forces}---forces on orbiting bodies arising from the emission, absorption or scattering of electromagnetic radiation---may induce slow changes to Keplerian orbits, in a similar manner as planet--planet interactions in multi-planet systems. The strength of radiative forces is proportional to the cross-sectional area of an orbiting body, while that of the gravitational force is proportional to its mass (thus, volume); therefore, radiative forces are most significant for the smallest bodies such as dust grains, and are unimportant for planetary bodies. These radiative forces become stronger as the stellar luminosity increases on the RGB and the AGB, and bodies for which these forces were previously insignificant begin to be affected by them.

\subsubsection{Radiation pressure}

\emph{Radiation pressure} is the radial force arising from momentum transfer from stellar photons to a body \citep{Burns79}. Its strength is given by
\begin{equation}
    F_\mathrm{rad}=\frac{L_\star A Q_\mathrm{PR}}{4\pi a^2 c},
\end{equation}
where $c$ is the speed of light, $A$ is the body's cross-sectional area and $Q_\mathrm{PR}$ its radiation pressure efficiency which includes absorption and scattering \citep{Burns79}. 
As the radiation pressure is a radial inverse square force, it can be considered as an effective reduction in stellar mass, with the parameter $\beta = F_\mathrm{rad}/F_\mathrm{grav}$, the ratio of radiative to gravitational force, being constant round the orbit. Objects on circular orbits that suddenly feel a strong radiation pressure (for example, dust grains released from an asteroid) will suddenly change their orbital eccentricity and semimajor axis, while the orbit will be unbound if $\beta \ge 0.5$ (this is the same mass reduction factor as leads to unbinding after impulsive mass loss described in Section~\ref{sec:2bodymassloss}). This condition is met at the \emph{blow-out size,} which is roughly $1\,\mu\mathrm{m}$ for dust orbiting the Sun. $F_\mathrm{rad}$ increases with increasing stellar luminosity, and the blow-out size increases to $\sim$mm sizes at the AGB tip. Particles below this limit are removed on an orbital timescale, as they follow hyperbolic trajectories, although it is worth noting that as the AGB tip luminosities are not maintained for long, such bodies may not actually be able to escape before the end of the AGB.

\subsubsection{Poynting--Robertson drag}

Let us now consider bound small particles, above the blow-out size. Such a small particle will absorb radiation from the star, be heated uniformly, and re-radiate the energy isotropically in its rest frame. When switching back to the inertial frame of the system, the emission therefore beamed in direction of orbital motion, resulting in a retarding force opposed to the orbital motion, known as \emph{Poynting--Robertson drag} or the \emph{Poynting--Robertson effect}. This creates a negative torque on the orbit leading to orbital in-spiral and eccentricity decay, with the rate of decay of semimajor axis being given by
\begin{equation}
\dot{a} = -\frac{L_\star Q_\mathrm{PR}A}{4\pi a m c^2}\frac{2+3e^2}{\left(1-e^2\right)^{3/2}},
\end{equation}
where $m$ is the body's mass \citep{Burns79}. In terms of the $\beta$ parameter introduced above, this leads to a lifetime of
\begin{equation}
    \frac{t_\mathrm{PR}}{\mathrm{yr}} = \frac{400}{\beta}\left(\frac{a}{\mathrm{au}}\right)^2,
\end{equation}
and so as $\beta$ increases significantly on the RGB and AGB orbital inspiral can become very rapid. 

\subsubsection{Yarkovsky effect}

Larger bodies in the km size range, instead of being uniformly heated and radiating isotropically, are large enough to have a significant temperature gradient across their volume, which can have seasonal and diurnal components \citep{Bottke06}. The result of this is an anisotropic absorption and emission of radiation, giving rise to a force acting to change the orbit, known as the \emph{Yarkovsky effect} or \emph{Yarkovsky force}. The orbit can expand or contract depending on the spin direction of the body. The Yarkovsky effect is responsible in the Solar System for delivering asteroids to unstable Kirkwood gaps in the Asteroid Belt. As radiation forces become significant on the AGB, rapid migration of bodies becomes possible.

\subsection{Physical alteration and destruction of bodies by radiation forces}

So far we have considered the effects of stellar evolution on bodies' orbits, which are always affected at least in their response to stellar mass loss. Stellar evolution can, in some cases, also have an effect on the bodies themselves, primarily through the star's increased luminosity. 

A simple estimate for the surface temperature of a body in orbit around a star is the \emph{equilibrium temperature} $T_\mathrm{eq}$, calculated by balancing the total incoming radiation from the illumination of one hemisphere by the star, and the total outgoing radiation assuming the body redistributes this energy effectively and reradiates it isotropically from its entire surface:
\begin{equation}
    T_\mathrm{eq} = \left[\frac{\left(1-A_\mathrm{b}\right)L_\star}{16\pi\epsilon\sigma a^2}\right]^{1/4}, \label{eq:Teq}
\end{equation}
where $A_\mathrm{b}$ is the Bond albedo of the body, $\epsilon$ its emissivity, and $\sigma$ the Stefan--Boltzmann constant. Despite the relatively weak $L_\star^{1/4}$ scaling, the strong increase of stellar luminosity entails a significant increase in bodies' equilibrium temperatures on the RGB and AGB.

\subsubsection{The YORP effect}

Related to the Yarkovsky effect is the \emph{Yarkovsky--O'Keefe--Radzievskii--Paddack (YORP)} effect, where anisotropic reflection, absorption and emission of starlight causes a torque on the body itself and affects its rotation rate and spin axis \citep{Bottke06}. The effect of this is seen in the Solar System, where small asteroids are spun up to the break-up limit. As do all radiation effects, while present on the MS it strengthens on RGB and AGB. Then bodies up to $\sim10$\,km can be rapidly spun up to break-up, leading to the release of their component parts (rocks, pebbles, etc) as debris \citep{Veras14}. Being much smaller than the parent body, these debris particles are then susceptible to rapid orbital evolution under the other radiative forces described above, and can migrate through the system to repopulate regions destabilised during a system's MS evolution and so provide a source for WD pollution.

\subsubsection{Sublimation and devolatilisation of bodies by strong irradiation}

Strong heating can result in physical alteration of bodies by sublimation of their more volatile components. Most easily lost are volatile species such as water with low sublimation points:for $1\mathrm{\,M}_\odot$ progenitors, smaller asteroids (few km) are totally dessicated out to 30--40\,au, while 100km asteroids are capable of retaining almost half of their water even at 5\,au \citep{Malamud16}. This process has implications for understanding extrasolar planetesimal compositions from WD pollution, as these have been devolatilised by RGB and AGB irradiation before we can observe them, and so may not reflect the composition of their progenitor bodies on the MS.

\subsubsection{Planetary atmospheric loss}

Planetary atmospheres are heated thermally as the star ascends RGB and AGB, when the stellar irradiation is intense but the photons of relatively low energy, since stars cool as they ascend giant branch. The heating of the planetary atmosphere, coupled with the accretion of stellar wind material for giant planets, can have effects on atmospheric chemistry \citep{Spiegel12}.

As the star and its expelled envelope transitions into the planetary nebula phase, the star core is briefly highly luminous in X-rays owing to its temperature. Planetary exospheres can then be heated by the XUV irradiation, leading to blow-off conditions and the loss of significant fractions of the planets' envelope. It is expected that the innermost giant planets that can survive stellar radius expansion and tidal interactions, within $\sim3$\,au, will lose most or all of their envelopes in this phase \citep{Villaver07}.

\subsubsection{Changes to planetary habitability}

The significant increase in stellar luminosity as the star ascends the RGB is ultimately lethal to Earth-like life on any planets in the liquid water habitable zone during the main sequence, as the planetary equilibrium temperature increases. Planets with surface water such as Earth will see a stronger increase in their surface temperature then suggested by Equation~\ref{eq:Teq} owing to the greenhouse effect, and Earth will likely become uninhabitable to complex life in roughly 1\,Gyr and to all life in around 3\,Gyr \citep{OMalleyJames13}, long before the Sun leaves the MS. As the Sun's luminosity continues to increase, Mars and the satellites of the giant planets will briefly enter the liquid water habitable zone. Whether life would have time to develop on analogous bodies in extra-Solar systems is an open question.

To end this section on a speculative note, a technological civilisation may attempt to construct a \emph{starshade,} a structure interposed between the planet and the star to counter the increasing stellar luminosity and maintain a habitable climate \citep{Gaidos17}. Even considering the relatively modest luminosity increase as the star progresses along the MS, these structures can be large enough to yield a transit signal detectable with current instrumentation, should they exist and cross our line of sight to the star.

\section{Summary and Conclusions}

We have seen that both stars and their planetary systems undergo significant changes after the star leaves the Main Sequence, progresses through RGB and AGB evolution, and becomes a white dwarf. Stellar mass loss causes orbits to expand, while the increased stellar radius enhances tidal forces which means that close-in planets out to a few au are engulfed into the envelope, where they likely do not survive unless they are around the brown dwarf mass limit ($\gtrsim10\mathrm{\,M_J}$). This means that a young white dwarf should be surrounded by a cleared volume, several au in radius, purged of all planetary material.

The fact that we see many white dwarfs surrounded by dusty debris discs within $\sim1\mathrm{\,R}_\odot$, or with rocky material in their atmospheres, together with a handful of transits of planets, or structures likely arising from disrupting asteroids, means that this picture cannot be complete. It is likely that stellar mass loss, in addition to causing orbits to expand, triggers orbital instability in a significant fraction of planetary systems orbiting WDs whose progenitors were stable on the main sequence. While in principle this is easy to understand, since after stellar mass moss the planets are relatively heavier compared to the star than they were before, the exact mechanism or mechanisms at play are not yet known, with changes to orbital resonances, changes to dynamics in stellar binary systems, and the outright destabilisation of entire planetary systems all likely playing some role in unknown proportions. Our understanding of the orbital dynamics at work here is hampered by our lack of knowledge of the relevant populations of planets: likely low-mass (sub-Saturn), on orbits of a few au or beyond (ensuring survival beyond the AGB), and orbiting stars of around $2\mathrm{\,M}_\odot$. These are all extremely challenging regimes to probe observationally.

Finally, the large stellar luminosity on the AGB also affects orbiting bodies. This can be through the enhancement of radiative forces such as radiation pressure, Poynting--Robertson drag, and the Yarkovsky and YORP effects, which can cause a significant orbital migration of dust and small asteroids; this may provide an additional source for delivery to white dwarfs. The large luminosity can also physically alter bodies, stripping planets of their atmospheres and asteroids and comets of their volatile content. These effects must also be taken into account when attempting to link planetary systems around white dwarfs to their progenitors around giant branch and main sequence stars, in order to get a full picture of the whole lifespans of planetary systems.

\begin{ack}[Acknowledgments]

AJM acknowledges support from the Swedish Research Council (Project Grant 2022-04043) and the Swedish National Space Agency (Career Grant 2023-00146). He would like to thank the Section Editor Dimitri Veras, and Paula Ronco, for useful comments on the manuscript.
\end{ack}

\seealso{An excellent and comprehensive review on the effects of stellar evolution on planetary systems, covering essentially all literature up to the time of its publication, can be found in \cite{Veras16}. Thorough treatments of celestial mechanics are to be found in \cite{MurrayDermott99} and \cite{Tremaine23}, while an introduction to the required stellar evolution theory is to be found in \cite{Prialnik09}.}

\bibliographystyle{Harvard}
\bibliography{reference}

\begin{thebibliography*}{37}
\providecommand{\bibtype}[1]{}
\providecommand{\natexlab}[1]{#1}
{\catcode`\|=0\catcode`\#=12\catcode`\@=11\catcode`\\=12
|immediate|write|@auxout{\expandafter\ifx\csname natexlab\endcsname\relax\gdef\natexlab#1{#1}\fi}}
\renewcommand{\url}[1]{{\tt #1}}
\providecommand{\urlprefix}{URL }
\expandafter\ifx\csname urlstyle\endcsname\relax
  \providecommand{\doi}[1]{doi:\discretionary{}{}{}#1}\else
  \providecommand{\doi}{doi:\discretionary{}{}{}\begingroup \urlstyle{rm}\Url}\fi
\providecommand{\bibinfo}[2]{#2}
\providecommand{\eprint}[2][]{\url{#2}}

\bibtype{Article}%
\bibitem[{Baronett} et al.(2022)]{Baronett+22}
\bibinfo{author}{{Baronett} SA}, \bibinfo{author}{{Ferich} N}, \bibinfo{author}{{Tamayo} D} and  \bibinfo{author}{{Steffen} JH} (\bibinfo{year}{2022}), \bibinfo{month}{Mar.}
\bibinfo{title}{{Stellar evolution and tidal dissipation in REBOUNDx}}.
\bibinfo{journal}{{\em Monthly Notices of the Royal Astronomical Society}} \bibinfo{volume}{510} (\bibinfo{number}{4}): \bibinfo{pages}{6001--6009}. \bibinfo{doi}{\doi{10.1093/mnras/stac043}}.
\eprint{2101.12277}.

\bibtype{Article}%
\bibitem[{Bonsor} et al.(2011)]{Bonsor11}
\bibinfo{author}{{Bonsor} A}, \bibinfo{author}{{Mustill} AJ} and  \bibinfo{author}{{Wyatt} MC} (\bibinfo{year}{2011}), \bibinfo{month}{Jun.}
\bibinfo{title}{{Dynamical effects of stellar mass-loss on a Kuiper-like belt}}.
\bibinfo{journal}{{\em Monthly Notices of the Royal Astronomical Society}} \bibinfo{volume}{414} (\bibinfo{number}{2}): \bibinfo{pages}{930--939}. \bibinfo{doi}{\doi{10.1111/j.1365-2966.2011.18524.x}}.
\eprint{1102.3185}.

\bibtype{Article}%
\bibitem[{Bottke} et al.(2006)]{Bottke06}
\bibinfo{author}{{Bottke} William~F. J}, \bibinfo{author}{{Vokrouhlick{\'y}} D}, \bibinfo{author}{{Rubincam} DP} and  \bibinfo{author}{{Nesvorn{\'y}} D} (\bibinfo{year}{2006}), \bibinfo{month}{May}.
\bibinfo{title}{{The Yarkovsky and Yorp Effects: Implications for Asteroid Dynamics}}.
\bibinfo{journal}{{\em Annual Review of Earth and Planetary Sciences}} \bibinfo{volume}{34}: \bibinfo{pages}{157--191}. \bibinfo{doi}{\doi{10.1146/annurev.earth.34.031405.125154}}.

\bibtype{Article}%
\bibitem[{Burns} et al.(1979)]{Burns79}
\bibinfo{author}{{Burns} JA}, \bibinfo{author}{{Lamy} PL} and  \bibinfo{author}{{Soter} S} (\bibinfo{year}{1979}), \bibinfo{month}{Oct.}
\bibinfo{title}{{Radiation forces on small particles in the solar system}}.
\bibinfo{journal}{{\em Icarus}} \bibinfo{volume}{40} (\bibinfo{number}{1}): \bibinfo{pages}{1--48}. \bibinfo{doi}{\doi{10.1016/0019-1035(79)90050-2}}.

\bibtype{Article}%
\bibitem[{Choi} et al.(2016)]{Choi16}
\bibinfo{author}{{Choi} J}, \bibinfo{author}{{Dotter} A}, \bibinfo{author}{{Conroy} C}, \bibinfo{author}{{Cantiello} M}, \bibinfo{author}{{Paxton} B} and  \bibinfo{author}{{Johnson} BD} (\bibinfo{year}{2016}), \bibinfo{month}{Jun.}
\bibinfo{title}{{Mesa Isochrones and Stellar Tracks (MIST). I. Solar-scaled Models}}.
\bibinfo{journal}{{\em The Astrophysical Journal}} \bibinfo{volume}{823} (\bibinfo{number}{2}), \bibinfo{eid}{102}. \bibinfo{doi}{\doi{10.3847/0004-637X/823/2/102}}.
\eprint{1604.08592}.

\bibtype{Article}%
\bibitem[{Collier Cameron} and {Jardine}(2018)]{CollierCameron18}
\bibinfo{author}{{Collier Cameron} A} and  \bibinfo{author}{{Jardine} M} (\bibinfo{year}{2018}), \bibinfo{month}{May}.
\bibinfo{title}{{Hierarchical Bayesian calibration of tidal orbit decay rates among hot Jupiters}}.
\bibinfo{journal}{{\em Monthly Notices of the Royal Astronomical Society}} \bibinfo{volume}{476} (\bibinfo{number}{2}): \bibinfo{pages}{2542--2555}. \bibinfo{doi}{\doi{10.1093/mnras/sty292}}.
\eprint{1801.10561}.

\bibtype{Article}%
\bibitem[{Debes} and {Sigurdsson}(2002)]{Debes02}
\bibinfo{author}{{Debes} JH} and  \bibinfo{author}{{Sigurdsson} S} (\bibinfo{year}{2002}), \bibinfo{month}{Jun.}
\bibinfo{title}{{Are There Unstable Planetary Systems around White Dwarfs?}}
\bibinfo{journal}{{\em The Astrophysical Journal}} \bibinfo{volume}{572} (\bibinfo{number}{1}): \bibinfo{pages}{556--565}. \bibinfo{doi}{\doi{10.1086/340291}}.
\eprint{astro-ph/0202273}.

\bibtype{Article}%
\bibitem[{Debes} et al.(2012)]{Debes12}
\bibinfo{author}{{Debes} JH}, \bibinfo{author}{{Walsh} KJ} and  \bibinfo{author}{{Stark} C} (\bibinfo{year}{2012}), \bibinfo{month}{Mar.}
\bibinfo{title}{{The Link between Planetary Systems, Dusty White Dwarfs, and Metal-polluted White Dwarfs}}.
\bibinfo{journal}{{\em The Astrophysical Journal}} \bibinfo{volume}{747} (\bibinfo{number}{2}), \bibinfo{eid}{148}. \bibinfo{doi}{\doi{10.1088/0004-637X/747/2/148}}.
\eprint{1201.0756}.

\bibtype{Article}%
\bibitem[{Gaidos}(2017)]{Gaidos17}
\bibinfo{author}{{Gaidos} E} (\bibinfo{year}{2017}), \bibinfo{month}{Aug.}
\bibinfo{title}{{Transit detection of a `starshade' at the inner lagrange point of an exoplanet}}.
\bibinfo{journal}{{\em Monthly Notices of the Royal Astronomical Society}} \bibinfo{volume}{469} (\bibinfo{number}{4}): \bibinfo{pages}{4455--4464}. \bibinfo{doi}{\doi{10.1093/mnras/stx1078}}.
\eprint{1705.01285}.

\bibtype{Article}%
\bibitem[{Gladman}(1993)]{Gladman93}
\bibinfo{author}{{Gladman} B} (\bibinfo{year}{1993}), \bibinfo{month}{Nov.}
\bibinfo{title}{{Dynamics of Systems of Two Close Planets}}.
\bibinfo{journal}{{\em Icarus}} \bibinfo{volume}{106} (\bibinfo{number}{1}): \bibinfo{pages}{247--263}. \bibinfo{doi}{\doi{10.1006/icar.1993.1169}}.

\bibtype{Article}%
\bibitem[{Lecar} et al.(2001)]{Lecar01}
\bibinfo{author}{{Lecar} M}, \bibinfo{author}{{Franklin} FA}, \bibinfo{author}{{Holman} MJ} and  \bibinfo{author}{{Murray} NJ} (\bibinfo{year}{2001}), \bibinfo{month}{Jan.}
\bibinfo{title}{{Chaos in the Solar System}}.
\bibinfo{journal}{{\em Annual Review of Astronomy and Astrophysics}} \bibinfo{volume}{39}: \bibinfo{pages}{581--631}. \bibinfo{doi}{\doi{10.1146/annurev.astro.39.1.581}}.
\eprint{astro-ph/0111600}.

\bibtype{Article}%
\bibitem[{Malamud} and {Perets}(2016)]{Malamud16}
\bibinfo{author}{{Malamud} U} and  \bibinfo{author}{{Perets} HB} (\bibinfo{year}{2016}), \bibinfo{month}{Dec.}
\bibinfo{title}{{Post-main Sequence Evolution of Icy Minor Planets: Implications for Water Retention and White Dwarf Pollution}}.
\bibinfo{journal}{{\em The Astrophysical Journal}} \bibinfo{volume}{832} (\bibinfo{number}{2}), \bibinfo{eid}{160}. \bibinfo{doi}{\doi{10.3847/0004-637X/832/2/160}}.
\eprint{1608.00593}.

\bibtype{Article}%
\bibitem[{Maldonado} et al.(2021)]{Maldonado21}
\bibinfo{author}{{Maldonado} RF}, \bibinfo{author}{{Villaver} E}, \bibinfo{author}{{Mustill} AJ}, \bibinfo{author}{{Ch{\'a}vez} M} and  \bibinfo{author}{{Bertone} E} (\bibinfo{year}{2021}), \bibinfo{month}{Jan.}
\bibinfo{title}{{Do instabilities in high-multiplicity systems explain the existence of close-in white dwarf planets?}}
\bibinfo{journal}{{\em Monthly Notices of the Royal Astronomical Society}} \bibinfo{volume}{501} (\bibinfo{number}{1}): \bibinfo{pages}{L43--L48}. \bibinfo{doi}{\doi{10.1093/mnrasl/slaa193}}.
\eprint{2010.11403}.

\bibtype{Book}%
\bibitem[{Murray} and {Dermott}(1999)]{MurrayDermott99}
\bibinfo{author}{{Murray} CD} and  \bibinfo{author}{{Dermott} SF} (\bibinfo{year}{1999}).
\bibinfo{title}{{Solar System Dynamics}}, \bibinfo{publisher}{{Cambridge University Press}}.
\bibinfo{doi}{\doi{10.1017/CBO9781139174817}}.

\bibtype{Article}%
\bibitem[{Mustill} and {Villaver}(2012)]{Mustill12}
\bibinfo{author}{{Mustill} AJ} and  \bibinfo{author}{{Villaver} E} (\bibinfo{year}{2012}), \bibinfo{month}{Dec.}
\bibinfo{title}{{Foretellings of Ragnar{\"o}k: World-engulfing Asymptotic Giants and the Inheritance of White Dwarfs}}.
\bibinfo{journal}{{\em The Astrophysical Journal}} \bibinfo{volume}{761} (\bibinfo{number}{2}), \bibinfo{eid}{121}. \bibinfo{doi}{\doi{10.1088/0004-637X/761/2/121}}.
\eprint{1210.0328}.

\bibtype{Article}%
\bibitem[{Mustill} et al.(2018)]{Mustill+18}
\bibinfo{author}{{Mustill} AJ}, \bibinfo{author}{{Villaver} E}, \bibinfo{author}{{Veras} D}, \bibinfo{author}{{G{\"a}nsicke} BT} and  \bibinfo{author}{{Bonsor} A} (\bibinfo{year}{2018}), \bibinfo{month}{May}.
\bibinfo{title}{{Unstable low-mass planetary systems as drivers of white dwarf pollution}}.
\bibinfo{journal}{{\em Monthly Notices of the Royal Astronomical Society}} \bibinfo{volume}{476} (\bibinfo{number}{3}): \bibinfo{pages}{3939--3955}. \bibinfo{doi}{\doi{10.1093/mnras/sty446}}.
\eprint{1711.02940}.

\bibtype{Article}%
\bibitem[{Naoz}(2016)]{Naoz16}
\bibinfo{author}{{Naoz} S} (\bibinfo{year}{2016}), \bibinfo{month}{Sep.}
\bibinfo{title}{{The Eccentric Kozai-Lidov Effect and Its Applications}}.
\bibinfo{journal}{{\em Annual Review of Astronomy and Astrophysics}} \bibinfo{volume}{54}: \bibinfo{pages}{441--489}. \bibinfo{doi}{\doi{10.1146/annurev-astro-081915-023315}}.
\eprint{1601.07175}.

\bibtype{Article}%
\bibitem[{Nordhaus} and {Spiegel}(2013)]{Nordhaus13}
\bibinfo{author}{{Nordhaus} J} and  \bibinfo{author}{{Spiegel} DS} (\bibinfo{year}{2013}), \bibinfo{month}{Jun.}
\bibinfo{title}{{On the orbits of low-mass companions to white dwarfs and the fates of the known exoplanets}}.
\bibinfo{journal}{{\em Monthly Notices of the Royal Astronomical Society}} \bibinfo{volume}{432} (\bibinfo{number}{1}): \bibinfo{pages}{500--505}. \bibinfo{doi}{\doi{10.1093/mnras/stt569}}.
\eprint{1211.1013}.

\bibtype{Article}%
\bibitem[{Nordhaus} et al.(2010)]{Nordhaus+10}
\bibinfo{author}{{Nordhaus} J}, \bibinfo{author}{{Spiegel} DS}, \bibinfo{author}{{Ibgui} L}, \bibinfo{author}{{Goodman} J} and  \bibinfo{author}{{Burrows} A} (\bibinfo{year}{2010}), \bibinfo{month}{Oct.}
\bibinfo{title}{{Tides and tidal engulfment in post-main-sequence binaries: period gaps for planets and brown dwarfs around white dwarfs}}.
\bibinfo{journal}{{\em Monthly Notices of the Royal Astronomical Society}} \bibinfo{volume}{408} (\bibinfo{number}{1}): \bibinfo{pages}{631--641}. \bibinfo{doi}{\doi{10.1111/j.1365-2966.2010.17155.x}}.
\eprint{1002.2216}.

\bibtype{Article}%
\bibitem[{O'Connor} et al.(2023)]{OConnor23}
\bibinfo{author}{{O'Connor} CE}, \bibinfo{author}{{Bildsten} L}, \bibinfo{author}{{Cantiello} M} and  \bibinfo{author}{{Lai} D} (\bibinfo{year}{2023}), \bibinfo{month}{Jun.}
\bibinfo{title}{{Giant Planet Engulfment by Evolved Giant Stars: Light Curves, Asteroseismology, and Survivability}}.
\bibinfo{journal}{{\em The Astrophysical Journal}} \bibinfo{volume}{950} (\bibinfo{number}{2}), \bibinfo{eid}{128}. \bibinfo{doi}{\doi{10.3847/1538-4357/acd2d4}}.
\eprint{2304.09882}.

\bibtype{Article}%
\bibitem[{O'Malley-James} et al.(2013)]{OMalleyJames13}
\bibinfo{author}{{O'Malley-James} JT}, \bibinfo{author}{{Greaves} JS}, \bibinfo{author}{{Raven} JA} and  \bibinfo{author}{{Cockell} CS} (\bibinfo{year}{2013}), \bibinfo{month}{Apr.}
\bibinfo{title}{{Swansong biospheres: refuges for life and novel microbial biospheres on terrestrial planets near the end of their habitable lifetimes}}.
\bibinfo{journal}{{\em International Journal of Astrobiology}} \bibinfo{volume}{12} (\bibinfo{number}{2}): \bibinfo{pages}{99--112}. \bibinfo{doi}{\doi{10.1017/S147355041200047X}}.
\eprint{1210.5721}.

\bibtype{Article}%
\bibitem[{Petit} et al.(2020)]{Petit20}
\bibinfo{author}{{Petit} AC}, \bibinfo{author}{{Pichierri} G}, \bibinfo{author}{{Davies} MB} and  \bibinfo{author}{{Johansen} A} (\bibinfo{year}{2020}), \bibinfo{month}{Sep.}
\bibinfo{title}{{The path to instability in compact multi-planetary systems}}.
\bibinfo{journal}{{\em Astronomy and Astrophysics}} \bibinfo{volume}{641}, \bibinfo{eid}{A176}. \bibinfo{doi}{\doi{10.1051/0004-6361/202038764}}.
\eprint{2006.14903}.

\bibtype{Article}%
\bibitem[{Petrovich} and {Mu{\~n}oz}(2017)]{Petrovich17}
\bibinfo{author}{{Petrovich} C} and  \bibinfo{author}{{Mu{\~n}oz} DJ} (\bibinfo{year}{2017}), \bibinfo{month}{Jan.}
\bibinfo{title}{{Planetary Engulfment as a Trigger for White Dwarf Pollution}}.
\bibinfo{journal}{{\em The Astrophysical Journal}} \bibinfo{volume}{834} (\bibinfo{number}{2}), \bibinfo{eid}{116}. \bibinfo{doi}{\doi{10.3847/1538-4357/834/2/116}}.
\eprint{1607.04891}.

\bibtype{Book}%
\bibitem[{Prialnik}(2009)]{Prialnik09}
\bibinfo{author}{{Prialnik} D} (\bibinfo{year}{2009}).
\bibinfo{title}{{An Introduction to the Theory of Stellar Structure and Evolution}}.

\bibtype{Article}%
\bibitem[{Reffert} et al.(2015)]{Reffert15}
\bibinfo{author}{{Reffert} S}, \bibinfo{author}{{Bergmann} C}, \bibinfo{author}{{Quirrenbach} A}, \bibinfo{author}{{Trifonov} T} and  \bibinfo{author}{{K{\"u}nstler} A} (\bibinfo{year}{2015}), \bibinfo{month}{Feb.}
\bibinfo{title}{{Precise radial velocities of giant stars. VII. Occurrence rate of giant extrasolar planets as a function of mass and metallicity}}.
\bibinfo{journal}{{\em Astronomy and Astrophysics}} \bibinfo{volume}{574}, \bibinfo{eid}{A116}. \bibinfo{doi}{\doi{10.1051/0004-6361/201322360}}.
\eprint{1412.4634}.

\bibtype{Article}%
\bibitem[{Ronco} et al.(2020)]{Ronco20}
\bibinfo{author}{{Ronco} MP}, \bibinfo{author}{{Schreiber} MR}, \bibinfo{author}{{Giuppone} CA}, \bibinfo{author}{{Veras} D}, \bibinfo{author}{{Cuadra} J} and  \bibinfo{author}{{Guilera} OM} (\bibinfo{year}{2020}), \bibinfo{month}{Jul.}
\bibinfo{title}{{How Jupiters Save or Destroy Inner Neptunes around Evolved Stars}}.
\bibinfo{journal}{{\em The Astrophysical Journal Letters}} \bibinfo{volume}{898} (\bibinfo{number}{1}), \bibinfo{eid}{L23}. \bibinfo{doi}{\doi{10.3847/2041-8213/aba35f}}.
\eprint{2007.04337}.

\bibtype{Article}%
\bibitem[{Sanderson} et al.(2022)]{Sanderson+22}
\bibinfo{author}{{Sanderson} H}, \bibinfo{author}{{Bonsor} A} and  \bibinfo{author}{{Mustill} A} (\bibinfo{year}{2022}), \bibinfo{month}{Dec.}
\bibinfo{title}{{Can Gaia find planets around white dwarfs?}}
\bibinfo{journal}{{\em Monthly Notices of the Royal Astronomical Society}} \bibinfo{volume}{517} (\bibinfo{number}{4}): \bibinfo{pages}{5835--5852}. \bibinfo{doi}{\doi{10.1093/mnras/stac2867}}.
\eprint{2206.02505}.

\bibtype{Article}%
\bibitem[{Smallwood} et al.(2021)]{Smallwood21}
\bibinfo{author}{{Smallwood} JL}, \bibinfo{author}{{Martin} RG}, \bibinfo{author}{{Livio} M} and  \bibinfo{author}{{Veras} D} (\bibinfo{year}{2021}), \bibinfo{month}{Jul.}
\bibinfo{title}{{On the role of resonances in polluting white dwarfs by asteroids}}.
\bibinfo{journal}{{\em Monthly Notices of the Royal Astronomical Society}} \bibinfo{volume}{504} (\bibinfo{number}{3}): \bibinfo{pages}{3375--3386}. \bibinfo{doi}{\doi{10.1093/mnras/stab1077}}.
\eprint{2104.06692}.

\bibtype{Article}%
\bibitem[{Spiegel} and {Madhusudhan}(2012)]{Spiegel12}
\bibinfo{author}{{Spiegel} DS} and  \bibinfo{author}{{Madhusudhan} N} (\bibinfo{year}{2012}), \bibinfo{month}{Sep.}
\bibinfo{title}{{Jupiter will Become a Hot Jupiter: Consequences of Post-main-sequence Stellar Evolution on Gas Giant Planets}}.
\bibinfo{journal}{{\em The Astrophysical Journal}} \bibinfo{volume}{756} (\bibinfo{number}{2}), \bibinfo{eid}{132}. \bibinfo{doi}{\doi{10.1088/0004-637X/756/2/132}}.
\eprint{1207.2770}.

\bibtype{Book}%
\bibitem[{Tremaine}(2023)]{Tremaine23}
\bibinfo{author}{{Tremaine} S} (\bibinfo{year}{2023}).
\bibinfo{title}{{Dynamics of Planetary Systems}}, \bibinfo{publisher}{{Princeton University Press}}.

\bibtype{Article}%
\bibitem[{Vassiliadis} and {Wood}(1993)]{Vassiliadis93}
\bibinfo{author}{{Vassiliadis} E} and  \bibinfo{author}{{Wood} PR} (\bibinfo{year}{1993}), \bibinfo{month}{Aug.}
\bibinfo{title}{{Evolution of Low- and Intermediate-Mass Stars to the End of the Asymptotic Giant Branch with Mass Loss}}.
\bibinfo{journal}{{\em The Astrophysical Journal}} \bibinfo{volume}{413}: \bibinfo{pages}{641}. \bibinfo{doi}{\doi{10.1086/173033}}.

\bibtype{Article}%
\bibitem[{Veras}(2016)]{Veras16}
\bibinfo{author}{{Veras} D} (\bibinfo{year}{2016}), \bibinfo{month}{Feb.}
\bibinfo{title}{{Post-main-sequence planetary system evolution}}.
\bibinfo{journal}{{\em Royal Society Open Science}} \bibinfo{volume}{3}, \bibinfo{eid}{150571}. \bibinfo{doi}{\doi{10.1098/rsos.150571}}.
\eprint{1601.05419}.

\bibtype{Article}%
\bibitem[{Veras} et al.(2011)]{Veras+11}
\bibinfo{author}{{Veras} D}, \bibinfo{author}{{Wyatt} MC}, \bibinfo{author}{{Mustill} AJ}, \bibinfo{author}{{Bonsor} A} and  \bibinfo{author}{{Eldridge} JJ} (\bibinfo{year}{2011}), \bibinfo{month}{Nov.}
\bibinfo{title}{{The great escape: how exoplanets and smaller bodies desert dying stars}}.
\bibinfo{journal}{{\em Monthly Notices of the Royal Astronomical Society}} \bibinfo{volume}{417} (\bibinfo{number}{3}): \bibinfo{pages}{2104--2123}. \bibinfo{doi}{\doi{10.1111/j.1365-2966.2011.19393.x}}.
\eprint{1107.1239}.

\bibtype{Article}%
\bibitem[{Veras} et al.(2013)]{Veras+13}
\bibinfo{author}{{Veras} D}, \bibinfo{author}{{Mustill} AJ}, \bibinfo{author}{{Bonsor} A} and  \bibinfo{author}{{Wyatt} MC} (\bibinfo{year}{2013}), \bibinfo{month}{May}.
\bibinfo{title}{{Simulations of two-planet systems through all phases of stellar evolution: implications for the instability boundary and white dwarf pollution}}.
\bibinfo{journal}{{\em Monthly Notices of the Royal Astronomical Society}} \bibinfo{volume}{431} (\bibinfo{number}{2}): \bibinfo{pages}{1686--1708}. \bibinfo{doi}{\doi{10.1093/mnras/stt289}}.
\eprint{1302.3615}.

\bibtype{Article}%
\bibitem[{Veras} et al.(2014)]{Veras14}
\bibinfo{author}{{Veras} D}, \bibinfo{author}{{Jacobson} SA} and  \bibinfo{author}{{G{\"a}nsicke} BT} (\bibinfo{year}{2014}), \bibinfo{month}{Dec.}
\bibinfo{title}{{Post-main-sequence debris from rotation-induced YORP break-up of small bodies}}.
\bibinfo{journal}{{\em Monthly Notices of the Royal Astronomical Society}} \bibinfo{volume}{445} (\bibinfo{number}{3}): \bibinfo{pages}{2794--2799}. \bibinfo{doi}{\doi{10.1093/mnras/stu1926}}.
\eprint{1409.4412}.

\bibtype{Article}%
\bibitem[{Villaver} and {Livio}(2007)]{Villaver07}
\bibinfo{author}{{Villaver} E} and  \bibinfo{author}{{Livio} M} (\bibinfo{year}{2007}), \bibinfo{month}{Jun.}
\bibinfo{title}{{Can Planets Survive Stellar Evolution?}}
\bibinfo{journal}{{\em The Astrophysical Journal}} \bibinfo{volume}{661} (\bibinfo{number}{2}): \bibinfo{pages}{1192--1201}. \bibinfo{doi}{\doi{10.1086/516746}}.
\eprint{astro-ph/0702724}.

\bibtype{Article}%
\bibitem[{Zahn}(1989)]{Zahn89}
\bibinfo{author}{{Zahn} JP} (\bibinfo{year}{1989}), \bibinfo{month}{Aug.}
\bibinfo{title}{{Tidal evolution of close binary stars. I - Revisiting the theory of the equilibrium tide}}.
\bibinfo{journal}{{\em Astronomy and Astrophysics}} \bibinfo{volume}{220} (\bibinfo{number}{1-2}): \bibinfo{pages}{112--116}.

\end{thebibliography*}

\end{document}